\documentclass{cpbtex}
\usepackage{graphicx}
\usepackage{bm}
\usepackage{color}
\begin{document}
\begin{CJK*}{GBK}{song}

\title{Numerical simulations of strong-field processes in momentum space \thanks{Project supported by the National Key Research and Development Program of China (Grant No. 2019YFA0307702), the National Natural Science Foundation of China (NSFC) (Grants Nos. 91850121 and 11674363), the Science Fund of Educational Department of Henan Province of China (Grant No. 2011C140001), and the Ninth Group of Key Disciplines in Henan Province (Grant No. 2018119).}}


\author{Yan Xu$^{1}$ and \ Xue-Bin Bian$^{2}$\thanks{Corresponding author. E-mail:xuebin.bian@wipm.ac.cn}\\
$^{1}${College of Physics and Electronic Engineering, Xinxiang University, Xinxiang 453003, China}\\  
$^{2}${State Key Laboratory of Magnetic Resonance and Atomic and Molecular Physics},\\
{Wuhan Institute of Physics and Mathematics,}\\
			{Innovation Academy for Precision Measurement Science and Technology,}\\
			{Chinese Academy of Sciences, Wuhan 430071, China}}


\date{\today}
\maketitle
\begin{abstract}
The time-dependent Schr\"{o}dinger equation (TDSE) is usually treated in the real space in the text book. However, it makes the numerical simulations of strong-field processes difficult due to the wide dispersion and fast oscillation of the electron wave packets under the interaction of intense laser fields. Here we demonstrate that the TDSE can be efficiently solved in the momentum space. The high-order harmonic generation and above-threshold ionization spectra obtained by numerical solutions of TDSE in momentum space agree well with previous studies in real space, but significantly reducing the computation cost.
\end{abstract}



\section{Introduction}\label{sec:I}
Strong-field physics has attracted a lot of attention in the last three decades. The nonperturbative phenomena, such as high-order harmonic generation (HHG), above-threshold ionization (ATI), double ionization (DI), above-threshold dissociation (ATD) \textit{et al}., have been well studied \cite{Corkum,Brabec}. For the nonlinear dynamic processes, there are seldom analytic solutions. Most of the theoretical works are based on numerical simulations of the time-dependent Schr\"{o}dinger equation (TDSE) \cite{Bian}. After obtaining the wavefunction at arbitrary times, all the experimental observables can be extracted from it.  However, it is not an easy task to calculate the time-dependent wavefunction precisely. The ionized wavepacket may be pushed very far away from the parent ion in real space. In addition, the kinetic energy of the photoelectron may be up to 10$U_p$ ($U_p$ is the ponderomotive energy.) after rescattering from the core \cite{Corkum,Brabec,Shi,Liu}. For example, if the laser intensity $I=1\times10^{14}$ W/cm$^2$, wavelength $\lambda=800$ nm, the full width at half maximum (FWHM) of the pulse $\sigma=$30 fs. The radial box in real space should be $R_{max}=2617$ a.u. to include the backscattering electrons. The density of the grids should be very high to accurately describe the fast oscillation of the high-energy wavefunction \cite{LYP}. At least, ten thousand of grids should be used. This is very time consuming during the evolution of the wavefunction, and the memory of the computer system should be big enough. To overcome this problem, parallel computing methods in real space have to be developed. Alternatively, a hybrid wavefunction splitting algorithm has been proposed\cite{Chelkowski}. In this scheme, the wavefunction near the atomic core is calculated in the real space. The external wavefunction is transformed into the momentum space by neglecting the long-range Coulomb potential to reduce the computation cost. In the split-operator method, the wavefunction is also transformed in different spaces by using FFT in every time step \cite{Feit}. In this work, different from previous schemes, we treat the whole TDSE and wavefunction directly in the momentum space \cite{William}, which can greatly reduce the cost of the numerical computations since both the bound and the continuum states are localized functions of the momentum $p$. The idea is similar to the treatment of period solids. They also have some differences. For periodic solids, the problem is transformed to the crystal momentum space based on the Bloch theorem, rather than the electron momentum space. The equations in one unit cell can be further treated in the momentum space by our method.

To our knowledge, this is the first work solving the strong-field problems of gases in the momentum space. The paper is organized as follows. We introduce the details of the numerical methods for solving TDSE in both real and momentum spaces in Sec. 2. The results and discussions of HHG and ATI spectra are presented in Sec. 3. We summarize our findings in Sec. 4.
\section{Theoretical methods}
\label{sec:II}

For simplicity, in this work we take the one-dimensional TDSE as an example to compare its solution in real and momentum spaces, respectively. Atomic units are used unless stated otherwise.

\subsection{TDSE in real space}

The TDSE with laser-atom interaction in the dipole approximation in real space is written as

\begin{equation}\label{E1}
i\dfrac{\partial \Psi(x,t)}{\partial t}=[-\frac{d^2}{2dx^2}+V(x)+xE(t)]\Psi(x,t).
\end{equation}

We use a model potential \cite{Zhou} $V(x)=-\dfrac{1}{\sqrt{1+x^2}}$ to represent the symmetric atomic system. The electric field of the laser pulse is obtained by

\begin{equation}\label{E2}
E(t)=-\dfrac{dA(t)}{dt},
\end{equation}
where $A(t)$ is the laser vector potential. In this work, its form is $A(t)=\frac{E_0}{\omega}\sin^2(\frac{t \pi}{\tau})\cos(\omega t)$, where $E_0$ is the field strength, $\omega$ is the angular frequency, and $\tau$ is the total pulse duration.

There are many numerical methods solving the eigenstates of the field-free system, such as $B$-spine basis expansion \cite{Zhou,Bian,Bian2}, spacial grid discretization \cite{Yuan,Feit,Bian3}, and so on. For the time evolution, there are Crank-Nicholson method \cite{Bian}, split-operator method \cite{Feit,Yuan}, Arnoldi/Lanczos method \cite{LYP,Peng}, and so forth. After obtaining the time-dependent wavefuction, the HHG spectra can be calculated by the Fourier transform of laser-induced dipole in the length, velocity and acceleration forms, respectively \cite{Bandrauk,Baggesen}. The box boundary $x_{max}$ should be bigger than $2E_0/\omega^2$ to include all the electrons being able to return to the parent ion at $x=0$ to emit high harmonics. Absorbing functions are utilized to avoid reflection from the boundary. For ATI, the boundary $x_{max}$ must be bigger than $\sqrt{20U_p}\tau$ to include all the rescattered electrons as discussed in Sec. 1. Otherwise, the information of high energy electrons will be lost by the absorbing function. The ATI spectra can be calculated from the final wavefunction after laser pulses by the energy window operator \cite{Muller,Schafer}, surface flux \cite{Yuan}, or projecting on the plane waves \cite{Madsen}, scattering states \cite{Zhou}, Coulomb waves \cite{LYP,Madsen}, and so on. Since there have been many studies, here we will not give the details.

\subsection{TDSE in momentum space}

The TDSE in the momentum space \cite{William} can be obtained by Fourier transform of Eq. (\ref{E1})

\begin{equation}\label{E3}
i\dfrac{\partial \Psi(p,t)}{\partial t}=\left\lbrace \frac{[p+A(t)]^2}{2}+\int_{-\infty}^{+\infty}\dfrac{V^\prime(p-p^\prime)}{2\pi}dp^\prime\right\rbrace \Psi(p,t).
\end{equation}

In this case, the term $V^\prime(p)$ is the Fourier transform of the Coulomb potential $V(x)$.

The eigenstates wavefunctions $|\psi_n(p)\rangle$ and eigenvalues $E_n$ are obtained by direct diagonalization of the field-free Hamiltonian matrix

\begin{equation}\label{E4}
H^0_{i,j}=\frac{p_i^2}{2}\delta_{ij}+\frac{dp}{2\pi}V^\prime(p_i-p_j).
\end{equation}

To calculate the time evolution of the wavefunction in Eq. (\ref{E3}), we adopt the split-operator method in the following form

\begin{align}
\Psi(p,t+\Delta t)=\exp(-iH^0\frac{\Delta t}{2}) \exp(-iH^I(t+\frac{\Delta t}{2})\Delta t)
\exp(-iH^0\frac{\Delta t}{2})\Psi(p,t)+O(\Delta t^3),  \nonumber	
\end{align}
where the laser-atom interaction term $H^I(t)=pA(t)+A^2(t)/2$. The evolution operator $\exp(-iH^0\Delta t/2)$ is expanded by $$\sum_n\exp(-i\delta tE_n/2)\lvert\psi_n(p)\rangle\langle\psi_n(p)\lvert$$ in its eigenstate space \cite{Tong}. This expansion is only calculated once before the time evolution. The evolution operator $ \exp(-iH^I\Delta t)$ can be directly applied to the wavefunction $\Psi(p,t)$ in the momentum space. So the time-evlolution process is very efficient.

After obtaining the wavefunction at arbitrary times, the transition dipole in the velocity form is

\begin{equation}\label{E5}
d_v(t)=\int_{-P_{max}}^{+P_{max}} \Psi^*(p,t)(\hat{p}+A(t))\Psi(p,t) dp.
\end{equation}

The HHG spectra are obtained by Fourier transform of $d_v(t)$

\begin{equation}\label{E6}
F(\Omega)=\left \lvert \int_0^\tau d_v(t)\exp(-i\Omega t)\right \lvert^2.
\end{equation}

To obtain the ATI spectra, we have to project out the bound states of system,

\begin{equation}\label{E7}
|\Psi^\prime(p,\tau)\rangle=|\Psi(p,\tau)\rangle-\sum_{n=1}^{nb}\alpha_n |\psi_n(p)\rangle,
\end{equation}
where $\alpha_n=\langle \psi_n(p)|\Psi(p,\tau)\rangle$ is the coefficient of the bound states, $nb$ is number of field-free bound states $|\psi_n(p)\rangle$. It is different from the cases in the real space, where the bound wave and ionized wave can be separated in space by additional evolution of TDSE without laser fields. However, in the momentum space, they can not be automatically separated. We have to project them out.

The momentum distribution of the photoelectron is the direct square of $|\Psi^\prime(p,\tau)\rangle$. The $+p$ wave and $-p$ wave correspond to the same ATI energy $E_e=p^2/2$. The ATI yield on the detector is the addition of the corresponding probability of wavefunction in the momentum space, which can be directly calculated by

\begin{equation}\label{E8}
Y(E_e=p^2/2)=\lvert \Psi^\prime(p_-,\tau) \lvert^2+\lvert \Psi^\prime(p_+,\tau) \lvert^2.
\end{equation}

There is no need to deal with the final wavefunction by different complex methods used for the case in real space as mentioned in the above subsection.

\section{Results and discussion}\label{sec:III}

In this section, we compare the eigenstates of the system and the HHG and ATI spectra calculated in different spaces.

\subsection{Eigenstate of the system}

For the field-free Hamiltonian in the momentum space in Eq. (\ref{E4}), the ground state can also be obtained by imaginary time propagation (ITP) \cite{Bian} by changing $t\rightarrow -it$. The converged ground state energy is $E_1=$-0.669775 a.u., agreeing well with the results by the ITP of Eq. (\ref{E1}) in real space. We show the ground state $\psi_1(x)$ in the real space in Fig. 1(a). The direct Fourier transform of $\psi_1(x)$ and the ground state  $\psi_1(p)$ in momentum space obtained by ITP of Eq. (\ref{E3}) are illustrated in Fig. 1(b).  One may find that they are in accord with each other. The distribution of the wavefunction in real space is around twice wider than that in momentum space. For a continuum state, the wavefunction oscillates in the whole real space, while the wavefunction is only a very narrow localized point in the momentum space. As a result, one may predict that the computation cost will be much smaller in the momentum space.

\begin{figure}
  \centering
  \includegraphics[width=9cm]{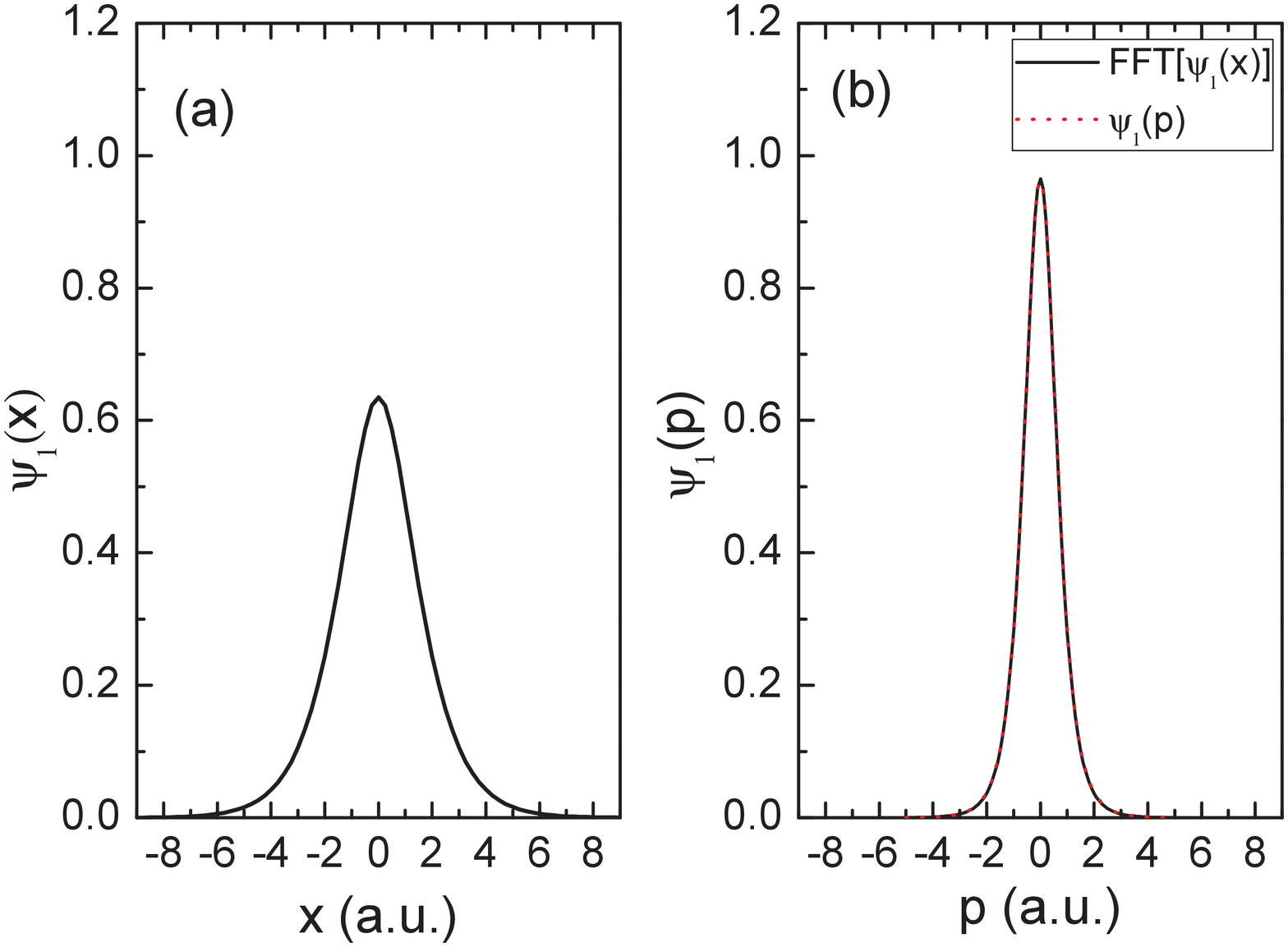}\label{Fig1}\\
  \parbox[c]{15.0cm}{\footnotesize{\bf Fig.~1.} Wavefunction of the ground state in the real space (a) and momentum space (b). The wavefunctions are normalized in the real space and momentum space, respectively. The direct Fourier transform of $\psi_1(x)$ in (a) is plotted by solid black line in (b).}

\end{figure}

\subsection{HHG and ATI spectra}

In this subsection, we calculate the HHG and ATI spectra of the model atomic system. The laser intensity we used is $I=1\times10^{14}$ W/cm$^2$. The wavelength $\lambda=800$ nm, the total pulse duration $\tau=10$ optical cycles. The HHG spectra calculated by the laser-induced dipole in the velocity form by solving the TDSE in the real space and momentum space are presented in Fig. 2. For clarity, we shift the red-dotted line down by three orders. One may find that the results are almost the same. In our simulations, the convergence has been checked. The range of $p$ is [-3, 3] a.u., $dp=0.01$ a.u., and the time step $dt=0.1$ a.u. The calculated cutoff energy agrees well with the law by the classical three-step model $I_p+3.17U_p=24\omega$, with $U_p=E_0^2/4\omega^2$ \cite{Corkum}.

\begin{center}

	\includegraphics[width=9cm]{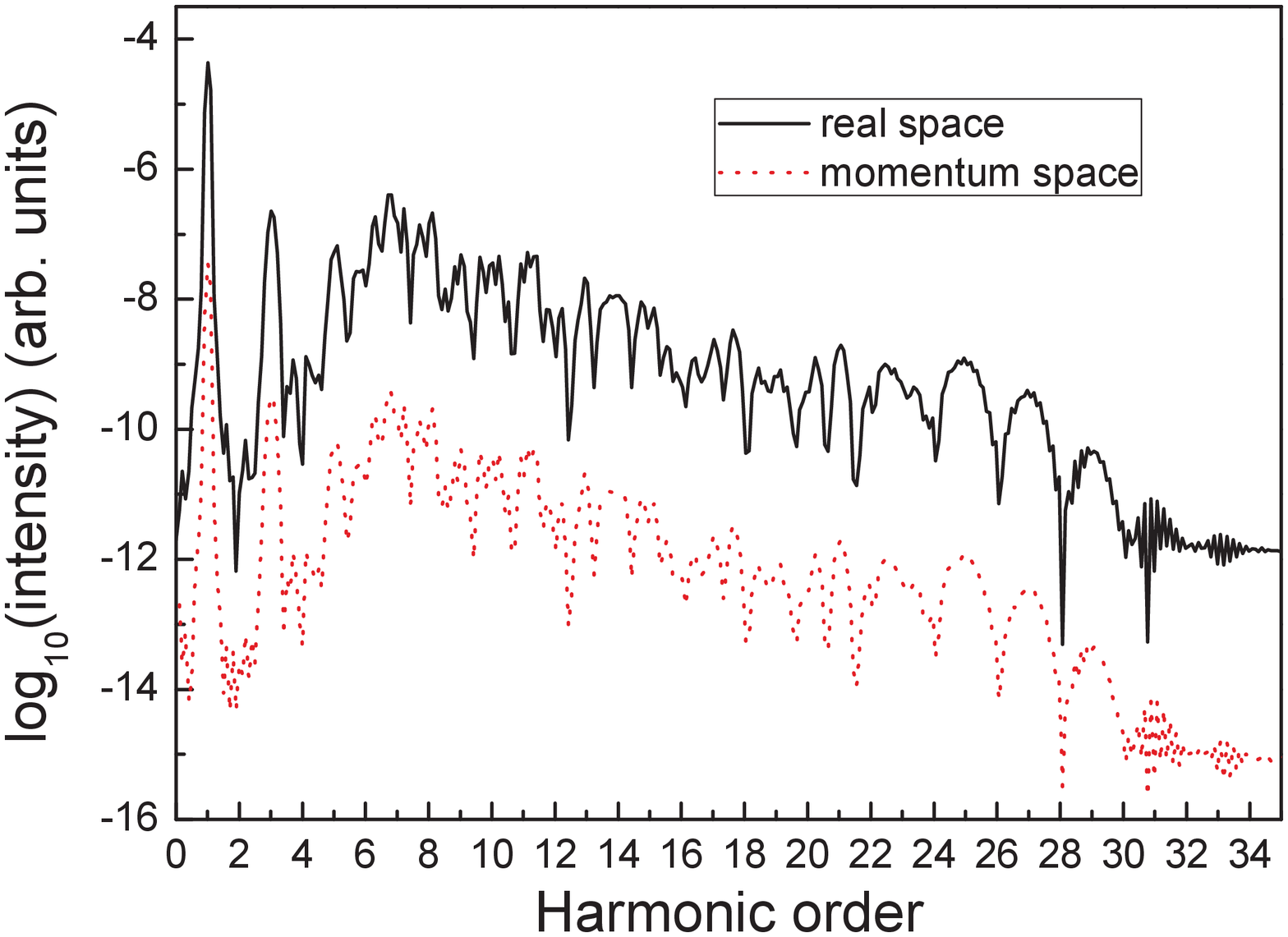}\label{Fig2}\\[5pt]  

	\parbox[c]{15.0cm}{\footnotesize{\bf Fig.~2.} Comparison of HHG spectra obtained by TDSE in the real and momentum spaces, respectively. For clarity, the red-dotted line is down shifted by three orders. The laser parameters can be found in the text.}

\end{center}

For the ATI spectra, it contains two main parts. One is the direct ionization electrons. Their maximum energy is around 2$U_p$ from the classical Newton equation. The other part is the rescattering electrons whose cutoff energy is around 10$U_p$ when the electron is backward rescattered. It is very hard to obtain the converged results in real space TDSE, especially near the 10$U_p$ cutoff. We use the method in Ref. \cite{Chelkowski} to treat the wavefunction by splitting algorithm to calibrate our results. The inner wavefunction is calculated by Eq. (\ref{E1}) in the real space, the outer wavefunction is calculated in the momentum space by neglecting the Coulomb potential. We compare the ATI spectra by the splitting algorithm and the direct evolution of TDSE in Eq. (\ref{E3}) in the momentum space in Fig. 3. One may find that the results agree well with each other. The positions of 2$U_p$ and 10$U_p$ cutoff energies are in accord with the predictions by classical simulations. In our calculations, the box in $p$ space is in a very small range of $[-4, 4]$ a.u., no matter how long the laser pulse is. For TDSE in real space, the box in the $x$ space is at least hundreds times larger to effectively retrieve the ATI yield in a femetosecond pulses. Longer pulses will further increase the box size and computation cost significantly in the real space. The resolution of energy is $dE=pdp$. For $10U_p$ electrons, the density of $p$ grids we used is $dp=0.0033$ a.u., which is around one order higher than that of $x$ grids. However, the total computation cost is still much lower in the momentum space even for a few-cycle short pulse.

\begin{center}

	\includegraphics[width=9cm]{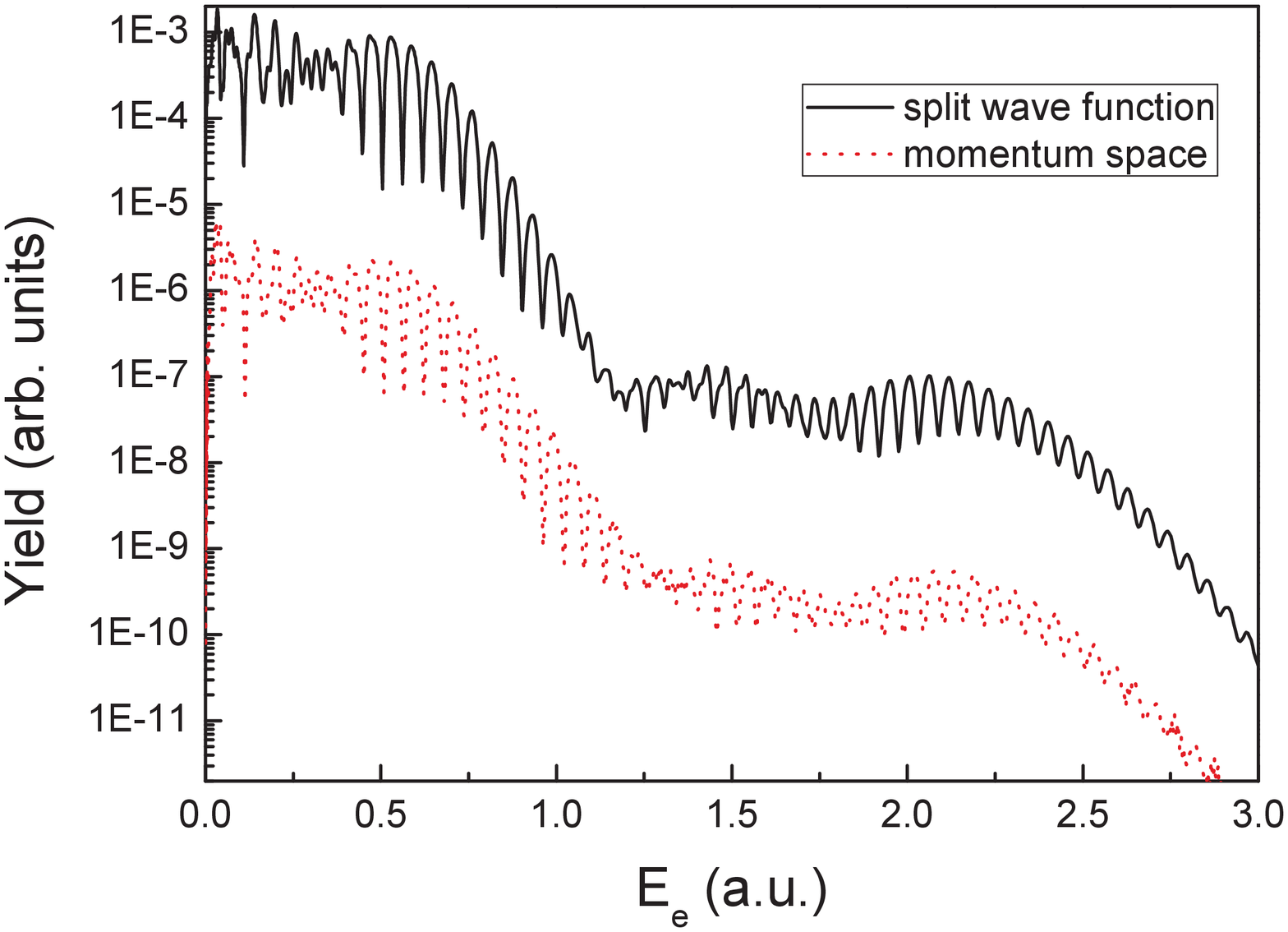}\label{Fig3}\\[5pt]  

	\parbox[c]{15.0cm}{\footnotesize{\bf Fig.~3.} Comparison of ATI spectral calculated by TDSE in the real and momentum spaces, respectively. The red dotted line is down shifted by two orders for clarity. The laser parameters are the same as those in Fig. 2.}

\end{center}

\section{Conclusions}\label{sec:IV}

In summary, we studied the eigenstates of the atomic Hamiltonian system in the momentum space. Both the bound and continuum states are localized waves in the momentum representation. We also solved the TDSE in the momentum space to simulate the strong-field HHG and ATI processes. Details of 1D calculations are illustrated. Good agreement has been achieved compared to the cases of real space. However, it reduces the computation cost greatly. It also provides us a new view of the strong-field processes in the momentum space. We will extend it to molecular system and higher dimensions in our future work. For the 3D TDSE in the spherical coordinates, it could be decomposed by the radial part and angular part with spherical harmonics in the momentum space. The range of the radial part in the momentum space should be small and reduce computation costs. In addition, it has been demonstrated that the velocity gauge is better than the length gauge in the simulation of ATI in real space by requiring a small number of spherical harmonics. Thus the 3D TDSE in the momentum space is expected to be efficient in the same way.

\end{CJK*}  
\end{document}